\documentclass[conference,hidelinks]{IEEEtran}
\usepackage{cite}
\usepackage{amsmath,amssymb,amsfonts}
\usepackage{algorithmic}
\usepackage{graphicx}
\usepackage{textcomp}
\usepackage[dvipsnames]{xcolor}
\usepackage{hyperref}
\usepackage{xspace}
\usepackage[T1]{fontenc}
\newcommand{\sol}{SOL\xspace}

\usepackage{letltxmacro}
\LetLtxMacro\oldttfamily\ttfamily
\DeclareRobustCommand{\ttfamily}{\oldttfamily\csname ttsize\endcsname}

\newcommand{\setttsize}[1]{\def\ttsize{#1}}%
\setttsize{\small} 

\usepackage{listings}
\usepackage{caption} 
\def\BibTeX{{\rm B\kern-.05em{\sc i\kern-.025em b}\kern-.08em
	T\kern-.1667em\lower.7ex\hbox{E}\kern-.125emX}}

\begin{document}

\title{\sol: Effortless Device Support for AI Frameworks without Source Code Changes}

\author{
\IEEEauthorblockN{Nicolas Weber and Felipe Huici}%
\IEEEauthorblockA{\textit{NEC Laboratories Europe}}%
}

\maketitle

\begin{abstract}
Modern high performance computing clusters heavily rely on accelerators to
overcome the limited compute power of CPUs. These supercomputers run various
applications from different domains such as simulations, numerical applications
or \textit{artificial intelligence} (AI). As a result, vendors need to be able
to efficiently run a wide variety of workloads on their hardware.

In the AI domain this is in particular exacerbated by the existance of a number
of popular frameworks (e.g, PyTorch, TensorFlow, etc.) that have no common code
base, and can vary in functionality. The code of these frameworks evolves
quickly, making it expensive to keep up with all changes and potentially forcing
developers to go through constant rounds of upstreaming.

In this paper we explore how to provide hardware support in AI frameworks
without changing the framework's source code in order to minimize maintenance
overhead. We introduce \sol, an AI acceleration middleware that provides a
hardware abstraction layer that allows us to transparently support heterogenous
hardware. As a proof of concept, we implemented \sol for PyTorch with three
backends: CPUs, GPUs and vector processors.
\end{abstract}

\begin{IEEEkeywords}
	artificial intelligence, middleware, high performance computing
\end{IEEEkeywords}

\section{Introduction}

\textit{Artificial Intelligence} (AI) has undoubtedly become one of the hottest
fields in computer science today, with software and hardware vendors alike
competing for a share of the big economic and scientific pie.

Within this domain, PyTorch and TensorFlow have surfaced as the most widely used
AI frameworks today\cite{2019_pytorch_vs_tf}, to the point that hardware vendors
are required to support at least one of these in order to get any kind of user
adoption. These frameworks are open source projects with large communities,
fastly evolving, requiring constant maintenance and code upstreaming. Because
this is a tedious and time consuming task that needs to be repeated for every
release of the framework, it has become common practise to branch the framework,
add hardware device support for it, and then publish the result as a separate
installation package\cite{rocm_tf, rocm_pytorch, tf_ve}. This makes the life of
the vendor much easier, but puts the maintenance burden on the user, who needs
to maintain different installations or Docker images of the same framework for
different devices. Worse, it prevents users from being able to mix devices from
different vendors in their AI applications.

In this paper we introduce \sol, a middleware that provides a hardware
abstraction layer that allows to simultaneously support different hardware
devices on a range of standard AI frameworks \emph{without} requiring changes to
the frameworks themselves. In fact, data scientists need only to add a few lines
of code to their scripts in order to enable \sol and its hardware support.

\begin{figure}[t]
	\resizebox{\linewidth}{!}{\includegraphics[]{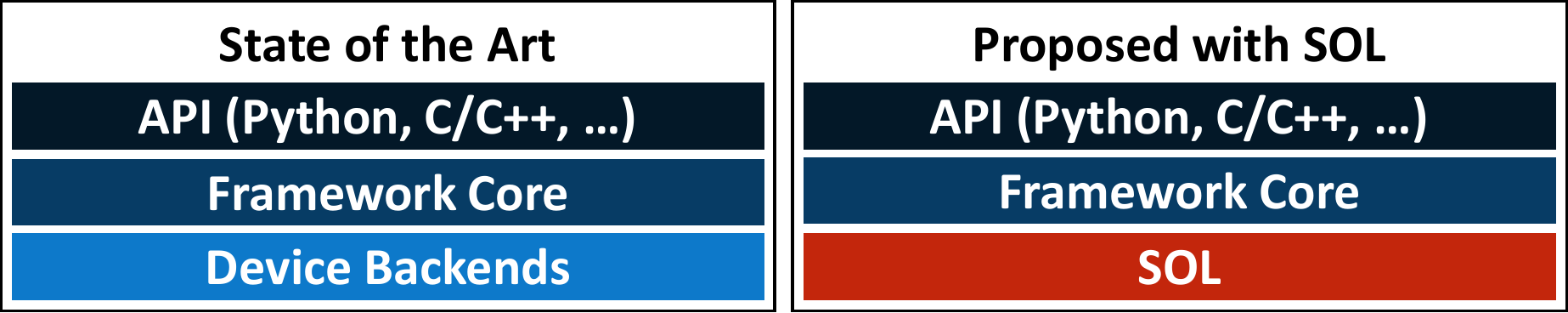}}
	\caption{Abstraction layers within AI frameworks.\vspace{-1em}}
	\label{fig_frameworks}
\end{figure}

We explore two strategies to integrate new devices into AI frameworks using \sol
as a middleware, to keep the original AI framework unchanged and still add
support to new device types. The first strategy hides the entire offloading
procedure from the framework, and the second only injects the necessary
functionality into the framework to enable the execution, but does so without
changing any PyTorch code. 

In short, our contributions are:

\begin{itemize}
  \item The design and implementation of \sol, a framework for
    \emph{transparently} adding heterogenous hardware device support to popular
    AI frameworks.
  \item The implementation and evaluation of two methods to integrate device
    support into AI frameworks without changing the framework's source code.
  \item An asynchronous execution design for device memory allocations.
  \item An evaluation of \sol on PyTorch using CPUs, GPUs and vector processor (NEC's SX-Aurora Tsubasa).
\end{itemize}

We show that \sol enables to add device support to AI frameworks with at max
3.000 lines of code per device backend and 2.400 lines per AI framework.
Further, because of the optimizing methods within \sol we achieve to accelerate
workloads up to (Inference/Training) 7.79x/2.41x (CPU), 4.37x/1.22x (GPU) and
25.41x/4.18x (NEC SX-Aurora) compared to the reference implementations within
the AI frameworks.

\section{Background on Neural Network Processing}
In this section we give an overview of existing AI frameworks, DNN optimization
libraries, and optimizing compilers/middleware.

\subsection{AI Frameworks}
The landscape of AI frameworks is rather vast, with Torch~\cite{torch} and
Theano~\cite{theano} being among the first popular frameworks. Facebook took
over the principles of Torch and switched its interface from Lua to Python,
introducing PyTorch~\cite{pytorch} and making it to one of the most widely used
frameworks today. PyTorch features a dynamic execution graph, making it very
flexible and easy to program. Google's TensorFlow~\cite{tf} is the biggest
competitor to PyTorch and uses a static execution graph (though it was recently
extended to also allow for dynamic graphs). There are also other frameworks such
as Chainer~\cite{chainer}, MxNet~\cite{mxnet}, and CNTK~\cite{cntk}, but their
user base is considerably smaller than those of PyTorch and
TensorFlow~\cite{2019_pytorch_vs_tf}.

What these frameworks have in common is their design of their internal
architecture (Fig.~\ref{fig_frameworks}). All of these frameworks have a frontend API, usually written in,
e.g., Python, C/C++ or Java, that maps onto a C-based core. This core handles
all of the framework's functionality. It manages the data on the devices and
processes the computation graphs of the neural networks by issuing function
calls to device specific backends. These backends either rely on hand optimized
compute kernels, written specifically for the framework, or on vendor specific
libraries, which we further explain in the next subsection.

The major problem for hardware vendors is, that there is no standard for these
device backends, making it necessary to write separate code for each framework.
DLPack~\cite{dlpack} is an approach for sharing data between deep learning
frameworks, but it did not get widely adopted. ONNX Runtime~\cite{onnxruntime}
and ONNXifi~\cite{onnxifi} aim to create device independent inference platforms
by keeping the device layer outside the framework. Regarding training, to the
best of our knowledge there is no existing work that abstracts the device layer
from the frameworks.

\subsection{Optimized DNN Libraries}
As NNs are very compute intensive, hardware vendors provide hand optimized
libraries to accelerate such workloads on their hardware: Intel provides
DNNL~\cite{dnnl} for their CPUs and GPUs, NVIDIA provides CUDNN~\cite{cudnn},
AMD provides ROCm MIOpen~\cite{miopen}, ARM provides ARMCL~\cite{armcl} and
NEC provides VEDNN~\cite{vednn}. These libraries all come with similiar
functionality, although ARMCL only supports functions for inference. Aside from
this, NNPACK~\cite{nnpack} provides functions for X86 and ARM64 CPUs, although
its performance is no longer competitive with respect to DNNL or ARMCL. All AI
frameworks usually rely on these libraries in order to leverage their superior
compute performance. Although the hyper parameters of NN layers are well defined, the APIs of these libraries are not, making it necessary to write separate code for each of these libraries.

\subsection{Optimizing Compilers and Middleware}
TVM~\cite{tvm} is an optimizing compiler architecture with a large open source
community and probably the highest number of supported hardware architectures.
It features integration for TensorFlow and PyTorch, but so far only supports
inference workloads (their high level IR ``Relay'' already supports training but
not the lower level implementations). One disadvantage of TVM for now is the
very long auto-tuning -- which can be up to several days, depending on the NN
structure and used hardware -- needed to reach good performance. Resnet-18 -- a
rather small network -- requires already 4h on an NVIDIA 1080
TI~\cite{tvm_resnet18}. Intel's OpenVino~\cite{openvino} is a similar tool,
targeting mainly Intel CPUs, Movidius VPUs or Intel GPUs. NVIDIA provides
TensorRT~\cite{tensorrt}, which is a similar but closed source tool, to deploy
NNs for NVIDIA hardware.

For PyTorch, AMD chose to write scripts that ``hiptify'' (rewrite) the PyTorch
source code, replacing all CUDA calls with HIP calls -- as they are
syntactically identical -- and then having their devices pose as CUDA ones
within PyTorch. However, this is only possible because they have spent
considerable effort mimicking the CUDA API and all of its libraries (i.e.,
CUBLAS, CUSOLVER, etc.) \cite{rocm}.

PlaidML~\cite{plaidml} and Tensor Comprehensions~\cite{tensorcomp} are both
compilers that use low level mathematical formulations rather than entire
layers, and transform these into specialized implementations for multiple
hardware architectures. These also require extensive auto-tuning to reach
performance comparable to hand-optimized libraries.

Brainslug~\cite{brainslug} was introduced by NEC as a framework to accelerate
workloads within AI frameworks using the depth first parallelism method. It
became the foundation of \sol's code optimization (see Sec.~\ref{sec_sol} for
more details). Aside from optimizing AI framework workloads, \sol also can
extract NNs from the frameworks and deploy in minimalistic libraries, removing
all framework dependencies, enabling to integrate these into user applications
for inference workloads. NGRAPH~\cite{ngraph} is a similar effort maintained by
Intel. It is mainly based on PlaidML to provide support for Intel CPU+GPU,
NVIDIA GPUs and AMD GPUs.

All of these approaches rely on similar optimizations ranging from high level
graph transformations, mathematical and algorithmic optimizations, down to low
level instruction optimizations that target specialized hardware through the use
of hand-optimized libraries for work-intensive operations.

In all, we are not aware of any work able to add heterogenous device support to
exisiting AI frameworks without changing their code base.

Beside these commonalities \sol targets some of the issues of modern AI
frameworks brought up by Barham and Isard~\cite{rut}. They criticize that
depending on the pass of the model execution, different memory layouts could be
optimal. \sol weights up if using the same layout in forward and backward pass
is faster than using separate layouts, including the necessary data
transformations. Further, \sol also allows to use different implementations for
forward and backward pass, i.e., to use OpenBLAS for forward and DNNL for
backward computations. Barham and Isard mention that AI frameworks address the
dimensions of a tensor by its numeric value. \sol instead uses identifiers
containing the purpose (None, Channel, Pixel) and a dimension. A tensor in NCHW
format has the dimensions \texttt{[N0, C0, P1, P0]} or \texttt{[N0, P1, P0, C0]}
in NHWC format. This enables \sol to make it easy to implement layers
independent of the used memory layouts, i.e., by automatically selecting all
channel dimensions for normalization layers, independent of how many there are.
On top, \sol's code generator automatically determines necessary nested loops
and how to map these onto the SIMD architecture of the hardware. More details
about this can be found in the following section.

\section{\sol Design and Implementation}
\label{sec_sol}

\begin{figure}[t]%
\vspace{-0.6em}
\noindent\begin{minipage}{0.986\linewidth}%
\begin{lstlisting}[caption={Full code example for using \sol within PyTorch. Line 5 optimizes the model, line 6 copies the parameters from \texttt{py\_model} to \texttt{sol\_model} and line 7 runs the optimized model.}, label={lst_code}, language=python]
import torch
import sol.pytorch as sol
py_model  = initPyTorchModel()
data      = initInputData()
sol_model = sol.optimize(py_model, data.size())
sol_model.load_state_dict(py_model.state_dict())
output    = sol_model(data)
\end{lstlisting}
\end{minipage}
\vspace{0.25em}
\resizebox{\linewidth}{!}{\includegraphics[]{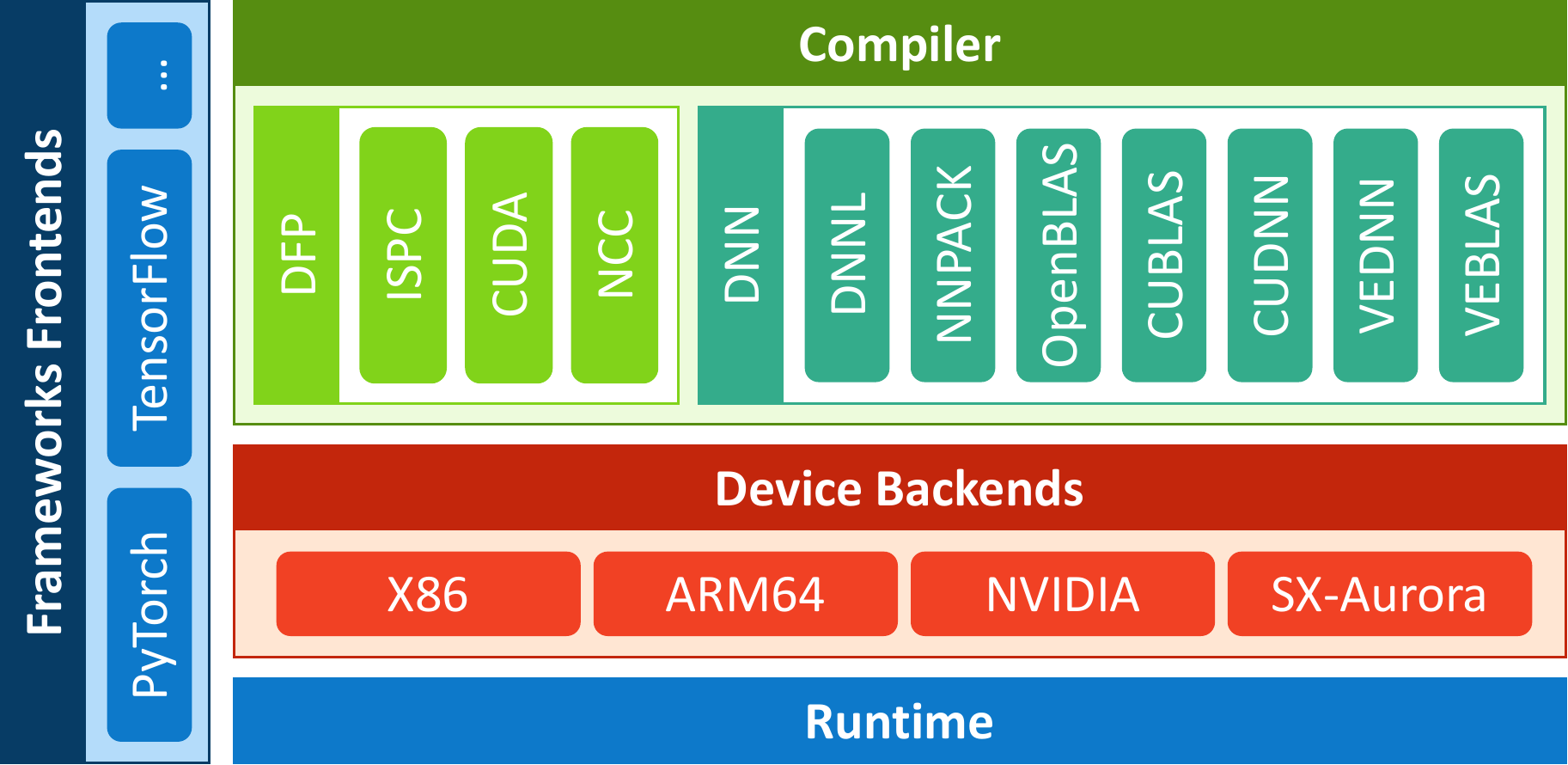}}
\caption{\sol architecture.\vspace{-1em}}
\label{fig_sol}
\end{figure}

\sol is an optimizing middleware for inference and training that transparently
integrates into existing AI frameworks (e.g., PyTorch or TensorFlow). It was
designed to have a very small programming footprint towards user; a data
scientist only needs to add a few lines of code to enable \sol (see
Listing~\ref{lst_code}).

Beyond transparency, \sol was explicitly designed to support multiple AI
frameworks (so called \emph{frontends}) and hardware devices (\emph{backends}),
including X86 and ARM64 CPUs, NVIDIA GPUs and the NEC SX-Aurora vector
processor. The core of it targets SIMD architectures in order to leverage common
features accross all of these hardware architectures within the same code base.
This allows us to write and maintain very small device backends (Figure~\ref{fig_sol} gives an overview of currently implemented modules).

In short, \sol consists of two main components, the compiler and the runtime. We
describe each of these in turn next.

\subsection{\sol Compiler}
The compiler is triggered by the call to \texttt{sol.optimize(...)}, which
extracts the computation graph from the framework and translates it into \sol's
own graph \textit{intermediate representation} (IR). First, \sol analyzes this
graph and applies general mathematic optimizations, i.e., a ReLU (\texttt{y =
max(x, 0)}) followed or preceeded by a MaxPooling can be removed from the graph
when the minimum value of the Pooling gets set to 0. In other cases the order of
layers can be switched without changing the mathematics, which can result in
better data reuse.

After these initial high level optimizations, the IR gets cloned for each device
type in order to apply device-specific optimizations. First, we determine which
optimizing method to apply to which layer. For now, we make this purely
heuristically, where all layers except Convolutions and Linears get implemented
using the \textit{Depth First Parallelism} (DFP) module~\cite{brainslug}. The
main idea of DFP is to process computation graphs in depth first order, to keep
data as long as possible in a processor's registers and caches; to achieve this
the DFP modules applies loop-transformation and fusion methods. The insight
behind the DFP principle is to generate code that minimizes the number of nested
loops while efficiently mapping these onto the SIMD architecture of the
hardware. The DFP module can handle arbitrary SIMD architectures from very short
AVX instructions to very long SX-Aurora vectors, and it is also able to make use
of features such as Shared Memory and SIMD-groups (warps).

Convolution and Linear layers get implemented with the \textit{Deep Neural
Network} (DNN) module, which maps these layers onto external libraries, such as
CUDNN or DNNL. There is one exception: if the Convolution is grouped and has as
many groups as output channels (e.g., in MobileNet) they get also implemented
using the DFP module, as this boils down to a WeightedPooling layer that can
make use of the depth first processing.

\sol further determines optimal memory layouts for the given data (e.g., DNNL
prefers blocked memory layouts) and takes care that data are always given in the
optimal layout to the layers, while trying to minimize the number of reorder
operations. For example, it turns out that for the Linear layer untransposed
weights (Output/Input Channels) work best for CPUs while (Input/Output Channels)
is faster on the NEC SX-Aurora. 

In case we have multiple libraries or algorithms or layouts available to
implement one of these layers, we either use heuristics or run a very short
auto-tuning workload to determine the best combination given the layer's
hyperparameters. \sol can mix the usage of different implementations, algorithms
and layouts between forward and backward pass to achieve higher performance.

After all layers have been assigned to an optimizing module, \sol generates code
for these and compiles it for the target devices. This entire optimization
procedure requires usually less than 1\,min (including the auto-tuning) and only
needs to be repeated if the input size of the network or its structure change.
After compilation, \sol injects a custom model into the AI framework so that the
user can use it the same way he would use a native model, with the difference
that the \sol model internally calls the optimized implementation when executed
(see Listing~\ref{lst_pymodel}).

\subsection{\sol Runtime}
The \sol runtime component connects the kernels with the framework's memory
allocation system; this makes it possible to directly read and write into the
framework's tensors without the need to copy between the framework's and \sol's
memory space. Further, AI frameworks usually pre-allocate device memory to speed
up allocations, which would limit the opportunity to maintain a separate memory
space. Further, the runtime component is responsible for loading the optimized
kernel functions, maintaining all communications between \sol, the framework and
the devices' APIs.

\begin{figure}[t]
\vspace{-0.375em}
\begin{lstlisting}[caption={Example of how \sol integrates its custom models into PyTorch.}, label={lst_pymodel}, language=python]
class SolModel(torch.nn.Module):
def __init__(self):
	self.param_0 = ...   # managed by framework
	self.param_1 = ...   # managed by framework

def forward(self, input):
	return sol.call(...) # executed by SOL
\end{lstlisting}
\end{figure}

\subsection{\sol Deployment}
Deployment is a special mode of the \sol compiler, that extracts the neural
network from AI frameworks to deploy it into a library that can be integrated
into a user application, similar to TVM, TensorRT or OpenVino. This specialized
NN library does not have any dependencies of the AI framework or \sol, only when
specialized functions are used from libraries such as DNNL or CUDNN.

\section{\sol Device Backends}
\sol device backends are very compact and easy to maintain. Each device relies
on one or multiple functional-backends that implement the functionality for the
DFP and DNN modules.

The DFP backends use a code generator that outputs standard C++ code. Only a few
function calls need to be overwritten to add device-specific ``flavours'' to the
generated code. Within the DFP generated code we use functions (i.e.,
\texttt{sol\_ispc\_exp}) that map onto device specific implementations. In most
cases this is just a \texttt{\#define sol\_ispc\_exp(A) exp(A)} but also can
contain specialized implementations, in case the device does not have specific
instructions for the given function. Listing~\ref{lst_dfp} shows how an
AveragePooling layer is described within the DFP module and how it is translated
into code for the different device backends.

The DNN backends only provide minimal functionality to initialize the libraries
descriptors and the ability to call the necessary library functions. The
descripters get initialized once when the neural network gets loaded and cached,
to decrease time during model execution. Further, the backends can implement
specialized auto-tuning functions to determine optimal algorithms or memory
layouts at compile time.

On top of these modules, the device backend can determine if the main thread
shall run on the host system or the device. 

\noindent\begin{minipage}{\linewidth}%
\begin{lstlisting}[caption={Example of an AveragePooling layer in DFP description and how it is translated to the different backends.}, label={lst_dfp}, tabsize=1]
// DFP: AveragePooling
auto I = layer(l->src()), O = layer(l);
auto K = kernel(l);
loop(); O += I[K];
loop(); O /= K.area(p->isCountPadding());
// Reference: standard C++
void kernel(const float* L0, float* L1) {
	for(int OC0x = 0; OC0x < 512; OC0x++)
	for(int OP1 = 0; OP1 < 128; OP1++)
	for(int OP0 = 0; OP0 < 128; OP0++) {
		float L1_s = 0;
		for(int K1 = 0; K1 < 3; K1++)
			for(int K2 = 0; K2 < 3; K2++)
				L1_s += L0[OC0x * 16384 + (OP1 + K1) * 128 + (OP0 + K0)];
		L1[OC0x * 16384 + OP1 * 128 + OP0] = L1_s / 9;
}}
// Backend-ISPC: X86 and ARM64
task void kernel(const uniform float* uniform L0, uniform float* uniform L1) {
	uniform int OC0x = taskIndex;
	foreach(OP1 = 0 ... 128, OP0 = 0 ... 128) {
		float L1_s = 0;
		for(uniform int K1 = 0; K1 < 3; K1++)
			for(uniform int K2 = 0; K2 < 3; K2++)
				L1_s += L0[OC0x * 16384 + (OP1 + K1) * 128 + (OP0 + K0)];
		L1[OC0x * 16384 + OP1 * 128 + OP0] = L1_s / 9;
}}
// Backend-CUDA: NVIDIA
__global__ void kernel(const float* L0, float* L1) {
	int OC0x = blockIdx.x;
	for(int OP0x = threadIdx.x; i < 16384; i += blockDim.x) {
		int OP1 = OP0x / 128, OP0 = OP0x % 128;
		float L1_s = 0;
		for(int K1 = 0; K1 < 3; K1++)
			for(int K2 = 0; K2 < 3; K2++)
				L1_s += L0[OC0x * 16384 + (OP1 + K1) * 128 + (OP0 + K0)];
		L1[OC0x * 16384 + OP0x] = L1_s / 9;
}}
// Backend-NCC: SX-Aurora
void kernel(const float* L0, float* L1) {
	#pragma omp parallel for collapse(2)
	for(int OC0x = 0; OX0x < 512; OC0x++) {
		#pragma _NEC ivdep
		for(int OP0x = 0; i < 16384; i++) {
			int OP1 = OP0x / 128, OP0 = OP0x % 128;
			float L1_s = 0;
			for(int K1 = 0; K1 < 3; K1++)
				for(int K2 = 0; K2 < 3; K2++)
					L1_s += L0[OC0x * 16384 + (OP1 + K1) * 128 + (OP0 + K0)];
			L1[OC0x * 16384 + OP0x] = L1_s / 9;
}}}
\end{lstlisting}
\end{minipage}

\noindent This can reduce communication overhead between host and device, if the devices
supports this implementation.

In the rest of the section we describe the implementation
of \sol's backends for CPU, GPU and the SX-Aurora.

\subsection{X86 and ARM64 Backend}
The backends for X86 and ARM64 both rely on the ISPC\cite{ispc} compiler for the
DFP generated code, as it allows to write very efficient vectorizable code. As
shown in Listing~\ref{lst_dfp} the syntax of ISPC varies from standard C++ by
keywords such as \texttt{uniform} (identifying a scalar variable) and
\texttt{foreach} (identifying a vectorized loop), but most of the structure is
identical to writing this in standard C++. For the DNN module, \sol's CPU
backends supports OpenBLAS, DNNL (only x86) and NNPACK.

\subsection{NVIDIA Backend}
The NVIDIA backend bears a close resemblance to the CPU one, except that it
relies on CUDA for the DFP module and CUBLAS and CUDNN for the DNN module. Again
Listing~\ref{lst_dfp} shows the differences. On top of the CPU backend, it
supports to use SIMD vector groups, which means that instead of using the full
vector length and sharing data using shared memory between the different warp,
\sol uses only the warp for vectorization. This allows to run multiple warps in
parallel, on different parts of the data, which improves performance in
situations where the available loops do not allow to leverage the combined SIMD
processing of all warps. For DNN we use the NVIDIA provided libraries CUDNN and
CUBLAS.

\subsection{SX-Aurora Backend}
The NEC SX-Aurora Tsubasa is a vector processor PCIe card. The SX-Aurora was not
specifically designed for NN processing, but for more traditional \textit{high
performance computing} (HPC) applications, such as numeric simulations. As a
result, it lacks AI-specific functionality such as tensor cores and float16
support. However, HPC clusters today need to run various kinds of workloads
including AI. To solve this, we implement an SX-Aurora \sol device backend to
transparently enable AI on this accelerator.

Developing the SX-Aurora backend for \sol was straightforward, as the
accompanying NCC compiler supports the C++14 standard and only requires very few
pragmas to be told which loop to vectorize. The DFP module is fairly slim, and
uses knowledge of vector lengths to ensure that vector instructions are not
underutilzed.

For the DNN module we use the VEDNN library~\cite{vednn} that was originally
implemented for the TensorFlow-VE~\cite{tf_ve} project. It containes optimized
implementations for Convolution, Linear, MaxPooling, SoftMax and Activation
layers, but we only use the Convolution and Linear implementations within \sol.
Additionally we use the SX-Aurora BLAS library as a secondary implementation for
Linear layers. For both libraries we use modified OpenMP implementations for
task parallelism, as the default implementations weakly scale.

As \sol and the AI frameworks are running on the host system, we use the
VEoffload~\cite{veoffload} library to launch our kernel functions on the Aurora.
It features a CUDA API-like programming model to asynchronously offload kernels
onto the device. However, it has latency issues because the execution queue is
operated by the host system. To address this, we build a specialized
asynchronous execution queue on top of the library specialized for the needs of
\sol. Our design mainly mimics the principles of CUDA streams, but extends it
with asynchronous malloc and free. As this does not directly allocate memory
immediately, we instead return a 64-bit integer, where the first 32 bits contain
a unique reference number and the second 32 bits can be used to offset the
pointer. This allows us to use this virtual pointer with normal pointer
arithmetics and removes the need to synchronize malloc and free operations,
increasing the asynchronity of the processing.

As a final optimization, we gather multiple adjacent memcopies and group them
together within our asynchronous execution queue. If only a small number of
small tensors need to be transferred, we use the latency- optimized VEoffload
memcopy methods. Otherwise, we use the peak bandwidth optimized
VEO-udma~\cite{veoffload_udma} library, which supports packed memcopies so that
many small tensors can be packed into a big data segment to speed up transfers.

\section{PyTorch Frontend}
To integrate \sol into AI frameworks we use so called \emph{frontends}. For the
PyTorch frontend, we developed two strategies: \textit{transparent offloading},
as it seamlessly integrates into the framework with only a minimal interaction
between PyTorch and the device backends; and \textit{native offloading} which
requires much tighter integration with the framework, but yields higher
performance during training.

\subsection{Transparent Offloading}
The idea behind transparent offloading is to add device support to the framework
with as minimal effort as possible. We took TensorFlow's Keras API as
inspiration, as it exposes two functions \texttt{model.predict()} and
\texttt{model.fit()} for prediction and training to the user, both of which
consume Numpy arrays located on the host system as input so the user does not
need to care about where the model is actually run. 

Normally when a model is executed, \sol checks on which device the input data is
located and executes all computations on that device. To enable transparent
offloading computations the user just needs to call
\texttt{sol.device.set(DEVICE, DEVICE\_IDX)} once prior executing the model. The
next time a model gets executed \sol recognizes that the data is located on the
host, but that the user requests to execute on another device. \sol then will
ensure that the necessary data is copied to the device and will execute the
model on it instead.

\sol injects its model into PyTorch's execution environment as a custom layer
(shown in Listing \ref{lst_pymodel}). This keeps the model parameters inside
PyTorch, so that \sol can leverage all available learning methods without the
need to implement these itself, and only execute the compute intensive
calculations within \sol.

One problem are the model parameters. As these are needed on the device, it
would be necessary to either copy them every time a model gets executed or to
cache these on the device. We chose the latter. When the model gets run for the
first time, we create a specialized offloading context that contains copies of
all model parameters. As long as the model parameters do not get modified or the
model gets destroyed, this context is kept alive to prevent continuous memcopies
between the host and the device, limiting memcopies between host and device to
just the input and output data.

While this works pretty well for inference, it is inefficient for training where
the model changes in each epoch; this means that we not only need to retransfer
the updated weights in each epoch but also to transfer all gradients from the
device to the host after the backward pass, as the gradient upgrade is processed
on the host system.

An obvious solution would be to implement the parameter management and learning
methods also within SOL. However, these features have different implementations
across AI frameworks, so portability is far from guranteed. As a result, we
decided to explore tighter integration with the framework.

\subsection{Native Offloading}
Support for X86, ARM64 and NVIDIA GPUs is already available in most AI frameworks, which allows \sol to directly connect to their public API to have access to all of the necessary framework functionality and to share the memory space on the device with the framework.

The SX-Aurora is not supported by these frameworks. AI frameworks such as
PyTorch and TensorFlow are built for multi-device support in mind and,
consequently, both of them support registering function callbacks for a specific
device. As these frameworks target extensibility, these callback registrations
are exposed to other libraries. We were interested to see if it would be
possible to integrate all the necessary operations into PyTorch without even
changing a single line of code, which would allow us to extend PyTorch without
initializing a tedious upstreaming and maintenance process. In principle we want
to be able to replace the entire lower device layer implementation of PyTorch
for the SX-Aurora with \sol. In the following we will reference source code
files which refer to the PyTorch source code located at
\href{https://github.com/pytorch/pytorch}{github.com/pytorch/pytorch}
for version 1.4.0.

First we analyzed how PyTorch distinguishes devices. Unfortunately they use a
fixed enum (\texttt{c10/core/DeviceType.h}), which cannot be extended from the
outside, so we decided to just take one of the existing devices (OpenCL, XLA or
HIP), as in the default installation package, only CPU and CUDA are used. The
\texttt{c10::RegisterOperators} class enables us to register the necessary
callbacks for devices within the device enum, as shown in
Listing~\ref{lst_register}.

\begin{figure}[t]
\vspace{-0.6em}%
\begin{lstlisting}[caption={Example to register a method \texttt{\_\_and\_\_} to PyTorch's callback registry.}, label={lst_register}]
at::Tensor __and__(const at::Tensor& A, const at::Tensor& B) { ... }
static auto registry = c10::RegisterOperators()
	.op(c10::RegisterOperators::options()
	.schema("aten::__and__.Tensor(Tensor self, Tensor other) -> Tensor")
	.kernel<at::Tensor(const at::Tensor&, const at::Tensor&)>(TENSOR_TYPE_ID, &__and__)
	.aliasAnalysis(c10::AliasAnalysisKind::FROM_SCHEMA))
\end{lstlisting}%
\vspace{-0.5em}%
\begin{lstlisting}[caption={PyTorch's DispatchStub that only supports CPU, CUDA and HIP functions.},label={lst_stub}]
template <typename rT, typename T, typename... Args>
struct DispatchStub<rT (*)(Args...), T> {
	// ...
	FnPtr cpu_dispatch_ptr;
	FnPtr cuda_dispatch_ptr;
	FnPtr hip_dispatch_ptr;
	// ...
};
\end{lstlisting}%
\vspace{-1em}%
\end{figure}

However, digging further through the source code reveals that some functions do
not get registered in the \texttt{c10::RegisterOperators} registry, but rely on
the class \texttt{at::native::DispatchStub} (\texttt{ATen/native/DispatchSub.h})
that only stores separate function pointers for CPU, CUDA and HIP
(Listing~\ref{lst_stub} shows an excerpt of that class). As CPU and CUDA are
already used within the default package, we chose to use HIP as our final device type.

Before the callbacks can be implemented, it is necessary to set up some basic
functionality, i.e., implementing the \texttt{at::HIPHooksInterface} which
contains methods to determine the number of available devices in the system, or
the default device index. Further, it is necessary to implement the
\texttt{at::Allocator} interface, which becomes the default allocator for the
given device. This is already enough to start implementing the first kernel
calls.

Our goal is to at least support the ability to print the contents of a tensor,
copy data between host and the device, and to run inference and training
workloads. For this it is necessary to implement a series of small functions to
create and reshape tensors, to fill tensors with scalar values and to read
scalar values from within the tensor. Further, we need some arithmetic kernels,
such as reductions (min, max, mean), unary (add, sub, div, mul, ...), logical
(lt, le, gt, ge, ...), operations to concatenate multiple tensors, binary (and,
or, ...) and finally calculations for the loss functions. This is sufficient to enable all of our required features.

\section{Evaluation}
We evaluate \sol under the aspect of how much effort it took to support the different processor architectures and how good performs \sol compared to the standard AI frameworks.

\subsection{Programming effort}
\sol was designed with extendibility and maintainability in mind. To add a new
layer to \sol only the layer description needs to be added to the high level IR
and to either implement it using the DFP or DNN module. All DFP layers are
automatically available for all devices due to the code generation engine, while
for the DNN module suitable glue code between \sol and the external libraries is
required.

Our X86 backend requires about 3.000 lines of code. For ARM64 we only
require 300 additional lines as it inherits most of its functionality from the
X86 backend. The NVIDIA GPU backend requires about 2.400 lines of code and the
NEC SX-Aurora about 2.200 lines of code, plus 800 lines dedicated to the kernels
required for the native tensor integration. We conclude that adding a device to
\sol requires at max 3.000 lines of code. In comparisson, we identified 26.000
lines for CPU and over 47.000 lines of code solely dedicated to NVIDIA GPUs
within PyTorch.


The frontend integration into PyTorch is about 1.200 lines of code for
extracting the neural network, injecting the \sol optimized model and to hook up
to the X86 and NVIDIA memory allocators to share the memory space with PyTorch.
The native tensor integration of the SX-Aurora required another 1.200 lines. In
total this is a rather small effort to do. It took a single programmer about 2
weeks to identify all entry points and implement the required callbacks and
kernel functions.

\subsection{Performance}
To evaluate the performance of \sol we ran tests for inference and for training
on an Intel Xeon 6126, an NEC SX-Aurora Tsubasa, an NVIDIA mid-range Quadro
P4000 and high-end Titan V (see Table~\ref{tbl_specs} for specs).

\begin{table}[t]
	\vspace{0.45em}
	\caption{Hardware devices used in our evaluation.}
	\label{tbl_specs}
	\resizebox{\linewidth}{!}{%
		\begin{tabular}{ |l|l|l|r|r|r| } 
		\hline
		Vendor	&	Model			& Type	& TFLOP/s	&	Bandwidth(GB/s)\\
		\hline
		Intel	&	Xeon Gold 6126	& CPU	& 0.88		&	119.21\\
		NEC		&	SX-Aurora VE10B	& VPU	& 4.30		&	1200.00\\
		NVIDIA	&	Quadro P4000	& GPU	& 5.30		&	243.30\\
		NVIDIA	&	Titan V			& GPU	& 14.90		&	651.30\\
		\hline
		\end{tabular}
	}
	\vspace{-1em}
\end{table}

We used PyTorch 1.4.0 as baseline for the CPU and GPUs. As the SX-Aurora is not
supported by PyTorch we used TensorFlow-VE 2.1 instead. The reference software
was installed using the official unmodified pip packages. In the following we
call the native execution model as \textit{SOL} and the transparent offloading
as \textit{SOL (TO)}.

As \sol currently supports CNN and MLP networks, we ran popular NN architectures
from these domains. While \sol has been tested with all models from the
TorchVision~\cite{torchvision} package due to space reasons we only report
results from Densenet, Resnet, Squeezenet, VGG, ShuffleNet v2, and MNasNet (two
versions each) and a 3-layer MLP with 8192 features and a ReLU activation
functions. ShuffleNet is not supported by TensorFlow-VE 2.1 as it does not
support 5D permutations. The CNN's input data is a tensor with the dimensions
$[B, 3, 224, 224]$ where $B$ stands for the batchsize. We repeated every
experiment 100 times.

\subsection{Inference}
\begin{figure*}[t]
	\resizebox{\linewidth}{!}{\includegraphics[]{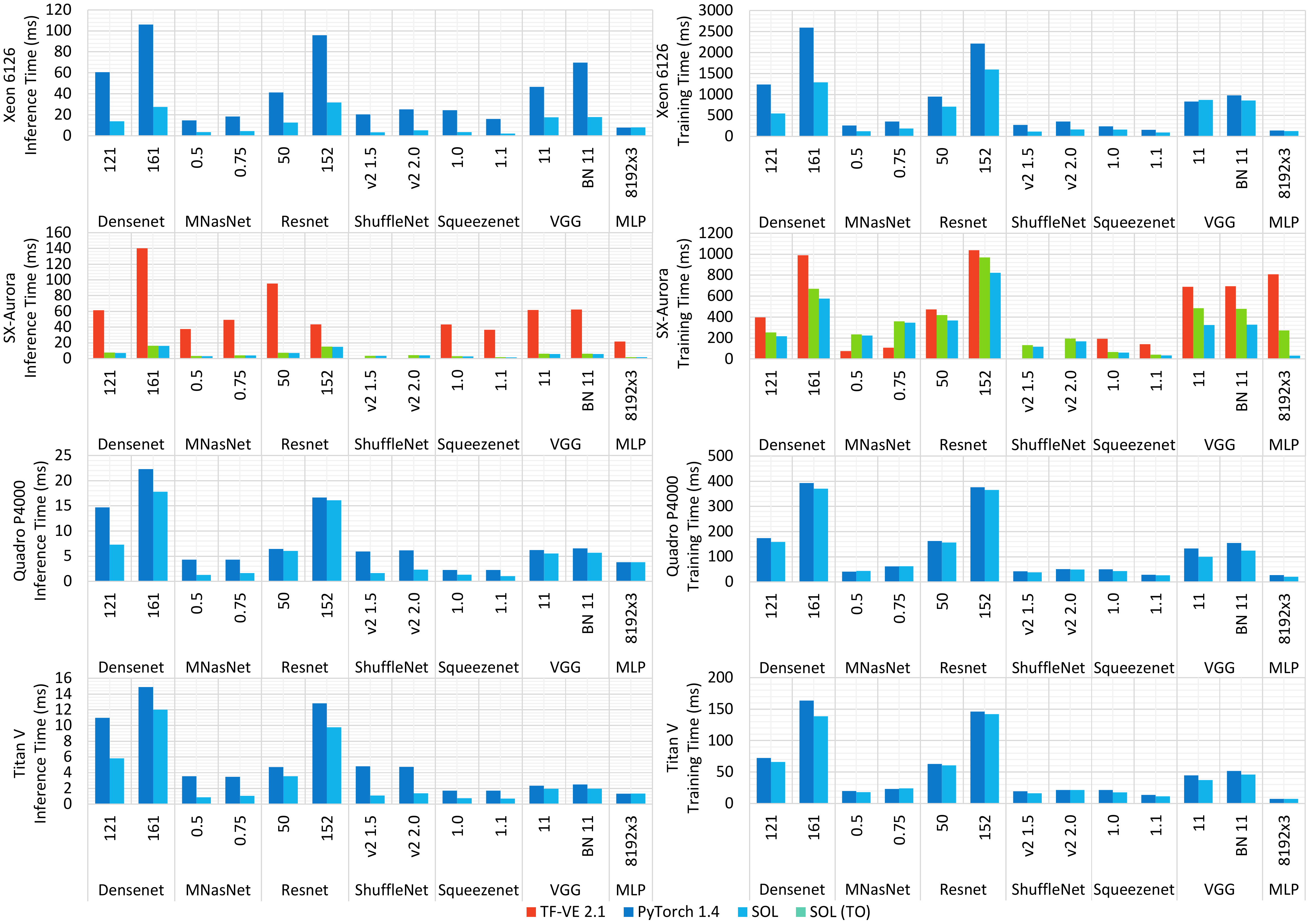}}
	\caption{\textbf{Left}: Inference ($B=1$) performance for all tested NNs and processors. \textbf{Right}: training ($B=16$ for CNNs and $B=64$ for MLP) performance. Results are reported as execution time in milliseconds.\vspace{-1.5em}}
	\label{fig_charts}
\end{figure*}

We start our performance evaluation with inference. The models are run with
$B=1$. For the CPU we can see that \sol is able to speed up the CNN models
significantly compared to the reference within PyTorch. For the MLP there is no
difference visible. MLPs do not provide optimization capabilities to \sol as it
mainly relies on matrix multiplications. In the SX-Aurora chart we can see that
TF-VE is always significantly slower than \sol. This is due to the VEDNN
library, that only parallelizes over the batch elements, so that only 1 out of 8
SX-Aurora cores is active. \sol uses a modified version of VEDNN with a
different, OpenMP-based parallization to overcome these problems. Further there
is no difference to be seen between the transparent and native offloading model,
as the data needed to be copied in inference is too small to make an actual
difference. In the GPU cases we can see, that \sol outperforms PyTorch
especially in DenseNet, Squeezenet and ShuffleNet. Overall \sol is always faster
than the baseline implementations in the inference tests, on all devices.

\subsection{Training}
For evaluating the CNNs we use $B=16$ and for the MLP $B=64$. The results are
shown in Fig.~\ref{fig_charts}. As before, for the CPU \sol is always faster,
especially in Densenet where the execution time is more than halfed. For the
SX-Aurora we see that TF-VE is always slowest except for the MNasNet. We
identified that \sol's code generated for the grouped convolutions is slower
than the implementation within VEDNN, which are used in TF-VE. For the other
networks SOL outperforms TF-VE with both execution modes, while as expected, the
native offloading always yields in higher performance, because of less memcopy
between the host and the device. The GPU performance gain of \sol is not as high
as for the inference cases, but still never slower than PyTorch.

\section{Conclusion}
In this paper we introduced \sol, an AI workload acceleration middleware that
makes it relatively easy to add support for a wide range of heterogenous
hardware devices to existing AI frameworks such as PyTorch and TensorFlow.
\sol's device backends add such support without having to modify the actual
framework's code, avoiding any upstreaming and maintanance hassles. As a proof
of concept, we implemented \sol along with a PyTorch frontend and backends for
CPU, GPU and a vector processor, and showed an extensive evalation of \sol
when running both inference and training workloads.

As future work we intend to add native tensor support for non-supported devices
to our other frontends (i.e. TensorFlow). In addition, we are looking into
supporting other kinds of networks such as transformers and RNNs.

\bibliographystyle{IEEEtran}
\bibliography{main}

\end{document}